\documentclass[preprint,twocolumn,10pt]{aastex6}
\usepackage{graphicx}
\usepackage{amsmath}
\usepackage{amssymb}
\usepackage{color}
\usepackage{cases}
\usepackage[bottom]{footmisc}
\usepackage{natbib}
\usepackage{soul}
\shortauthors{Grimaldi et al.}
\shorttitle{Area Coverage of E.T. Signals}

\begin{document}
\pagenumbering{arabic}

\title{Area Coverage of Expanding E.T. Signals in the Galaxy: SETI and Drake's $N$ }

\author{Claudio Grimaldi\altaffilmark{1}, Geoffrey W. Marcy\altaffilmark{2}, Nathaniel~K.~Tellis\altaffilmark{3}, Frank Drake\altaffilmark{4}}
\email{claudio.grimaldi@epfl.ch}
\email{geoff.w.marcy@gmail.com}
\altaffiltext{1}{Laboratory of Physics of Complex Matter, Ecole Polytechnique F\'ed\'erale
de Lausanne, Station 3, CP-1015 Lausanne, Switzerland}
\altaffiltext{2}{Professor Emeritus, University of California, Berkeley, CA 94720, USA} 
\altaffiltext{3}{Department of Astronomy, University of California, Berkeley, CA 94720, USA}  
\altaffiltext{4}{SETI Institute; Professor Emeritus, University of California, Santa Cruz, CA}

\begin{abstract}
The Milky Way Galaxy contains an unknown number, $N$, of civilizations that emit 
electromagnetic radiation (of unknown wavelengths) over a finite lifetime, $L$. 
Here we are assuming that the radiation is not produced indefinitely, but within $L$ 
as a result of some unknown limiting event. When a civilization stops emitting, 
the radiation continues traveling outward at the speed of light, $c$, but is confined 
within a shell wall having constant thickness, $cL$. We develop a simple model of 
the Galaxy that includes both the birthrate and detectable lifetime of civilizations 
to compute the possibility of a SETI detection at the Earth. Two cases emerge 
for radiation shells that are (1) thinner than or (2) thicker than the size of the 
Galaxy, corresponding to detectable lifetimes, $L$, less than or greater than the 
light-travel time, $\sim 100,000$ years, across the Milky Way, respectively. For case (1), 
each shell wall has a thickness smaller than the size of the Galaxy and intersects
the galactic plane in a donut shape (annulus) that fills only a fraction of the 
Galaxy's volume, inhibiting SETI detection. But the ensemble of such shell walls
may still fill our Galaxy, and indeed may overlap locally, given a sufficiently high
birthrate of detectable civilizations. In the second case, each radiation shell is
thicker than the size of our Galaxy. Yet, the ensemble of walls may or may not
yield a SETI detection depending on the civilization birthrate. We compare the 
number of different electromagnetic transmissions arriving at Earth to Drake's $N$,
the number of currently emitting civilizations, showing that they are equal to 
each other for both cases (1) and (2). However, for $L < 100,000$ years, the 
transmissions arriving at Earth may come from distant civilizations long extinct,
while civilizations still alive are sending signals yet to arrive.

\end{abstract}

\maketitle

\section{Introduction}
\label{intro}

Strategies to search for other technological civilizations in the Milky Way Galaxy or in other galaxies (SETI) often employ concepts 
from the Drake equation \citep{Drake1961, Drake2010} to parametrize the number and density of civilizations in the Milky Way Galaxy 
and its sub-components.  Among the key terms is the fraction of typical stars in the Milky Way (spectral types F, G, K, or M) that harbor 
Earth-size planets orbiting far enough from their host star that their surface temperature enables water to be liquid. 
Analysis of the data from the NASA Kepler mission \citep{Borucki2010} shows that fraction to be 10-20\% \citep{Petigura2013} for stars 
roughly similar to the Sun (FGK type).
For smaller stars, the M dwarfs, the occurrence may be even higher \citep{Dressing2013}. The implied number of warm 
($0$-$100$ $^\circ$C surface temperature), Earth-size planets in the Milky Way is roughly $20$-$40$ billion. Whether they indeed have 
and retain liquid water during Darwinian time scales remains unknown.

A key factor in the Drake equation is $L$, the typical lifetime of detectable technological civilizations, defined 
to be those capable of emitting electromagnetic (EM) radiation (or particles), with an intensity revealing their presence 
to others within the Milky Way. Various estimates of $L$ have been made, including $12,100$ years \citep{Gott1993} and $106$ years \citep{Cameron1963}, 
based either on a probabilistic argument of our human current terrestrial lifetime experience or 
on observed lifetimes of other species. Of course, Homo sapiens face unique capabilities of self-destruction, as well as preservation via 
planetary diversity, rendering our lifetime unknown by orders of magnitude. The computed value of the number of civilizations, $N$, remains 
uncertain by many orders of magnitude, primarily because we don't know the average value of $L$ or the fraction 
of warm Earth-size planets that spawn technological life.

The prevalence of Earth-size planets has sparked a resurgence in SETI efforts \citep{Siemion2015}, fueled by the Breakthrough Prize Foundation 
and its initiative, ``Listen'' \citep{Isaacson2017, MacMahon2017}. Listen uses both large radio telescopes outfitted with broadband radio 
frequency analyzers and optical telescopes outfitted with high resolution spectrometers, to search stars and galaxies (and other exotica) for 
EM waves that are inconsistent with natural emission mechanisms, but could be of technological origin, notably confined 
to a narrow range of frequencies and/or times (in the case of pulsed transmissions).

The value of $N$ in the Drake equation would seem to offer a framework to optimize SETI strategies and estimate the probability of 
success \citep{Tarter2001}. If $N$ were greater than $1$ in the Milky Way, SETI efforts would seem compelling. However the value of Drake's 
$N$ does not account for the limited and confined volume of space occupied by the EM waves from a civilization. With their limited 
lifetimes, civilizations fill a limited volume of space with their EM waves, depending on $L$. For lifetimes, $L<100,000$ years, 
the volume filled by their EM waves will be smaller than the volume of the Galaxy, decreasing the probability of detection. 
For comparison, humans have been transmitting radio waves into the Galaxy steadily for $80$ years, filling a bubble that has a volume less 
than $0.001$ \% of the volume of the Milky Way Galaxy.  SETI programs located throughout the Galaxy would have a low probability of detecting 
our signals. Thus, Drake's $N$ does not represent the number of electromagnetic signals hitting Earth from the galactic 
civilizations that are currently transmitting if their $L$ is less than $100,000$ years.

Here we attempt to bridge the gap between Drake's $N$ and the limited volume in the Galaxy filled by ETI's electromagnetic transmissions, 
whose detectable transmissions have been terminated, toward informing SETI efforts. We proceed by constructing a simple 
model of the Galaxy containing technological civilizations that form at a specified rate and emit EM waves for a specified time. 
The waves propagate through the Galaxy, whether the civilization is still emitting or not, filling volumes that we compute and track. 
This model is similar to that presented in \citet{Grimaldi2017}. Here we explicitly include the geometry of the emissions, the birthrate 
of civilizations, and the relationship to Drake's $N$.  
A video constructed in 2015 illustrates the basic geometrical concept of expanding bubbles of EM emission that become empty 
when the transmission ceases, filling only a fraction of the Galaxy.\footnote{\url{http://w.astro.berkeley.edu/~gmarcy/MW_SETI.mp4}} 
We aim to compute the number of electromagnetic signals washing over the Earth and compare it to $N$ of the Drake 
Equation \citep{Drake1961, Drake1992, Drake2010}.

\section{Geometrical Development}
\label{geo}
\subsection{Overarching Geometry}
\label{over}

We consider the Milky Way Galaxy with its roughly $200$ billion stars, located within a radius of approximately $R_G = 50,000$ lyr 
from the galactic center, containing tens of billions of potentially habitable planets \citep{Petigura2013}.  The Earth resides 
$\sim 27,000$ lyr  ($8.3\pm 0.3$ kpc) from the galactic center.  We adhere to the level of approximation of the Drake equation,
befitting our considerable ignorance about the biological origins of single-cell and intelligent life, and ignoring stellar density 
and metallicity gradients within the Galaxy as well as the shapes of its subcomponents. 

We consider technological civilizations that pop into existence at various locations within the Galaxy.  We presume that these civilizations 
are ``born'' (begin emitting EM waves) within the Milky Way at some rate,  $\eta_b$. That such a birthrate may be defined with 
an associated time interval between births has been discussed by \citet{Carter1983} and \citet{Gott1993} in terms of convergent evolution 
and the coincidence of time scales between the remaining main sequence lifetime of the Sun, the age of the Earth, and the evolutionary time 
scale required for the emergence of humans as a representative technological species. Such reasoning leads to a crude estimate of the time 
interval for emergence of technological civilization of several billion years per habitable planet, as on Earth. This time scale leads to a 
crude estimate of $\eta_b\lesssim 0.01$ per year in the Galaxy \citep{Gott1993} if the evolution of life on Earth offers a rough representative 
guide. For example, we may imagine (somewhat more pessimistically) that within the Milky Way civilizations are born at a rate of one per $1000$ 
years, yielding $\eta_b= 0.001$ yr$^{-1}$. We define $\tau_b\equiv \eta_b^{ -1}$ to be the typical time between births, in this 
example $1000$ years.

Here, we adopt the presumption that $\eta_b$  is independent of $L$.  Such would not be the case if new independent transmitters 
were spawned by civilizations themselves, or by civilizations that engage in interstellar colonization, effectively reproducing. 
In the former case, each civilization effectively gives birth to some number of EM transmitters, yielding a multiplicative factor 
that increases $\eta_b$ . In the latter case, civilizations would venture and proliferate exponentially, yielding growth modeled by 
standard population dynamics, and eventually filling the Galaxy. Such seems not to be the case in the Milky Way, but the possibility 
remains. Here we adopt the simple assumption that habitable planets spawn technological civilizations after a certain time of Darwinian 
gestation, leading to independence of $L$ and $\eta_b$.

\begin{figure}[t]
\begin{center}
\includegraphics[scale=0.3,clip=true]{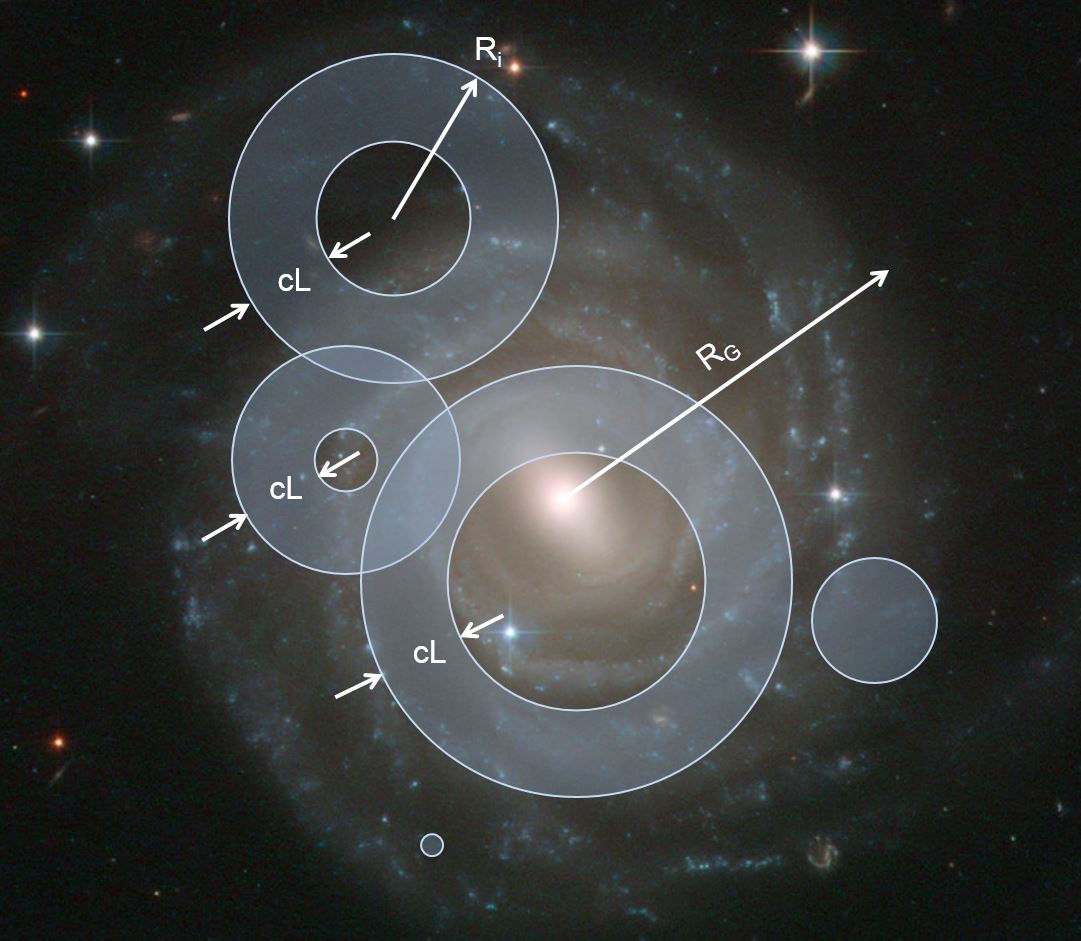}
\caption{A schematic of the Milky Way Galaxy showing the annuli of EM radiation emitted by five technological civilizations that formed, 
three of which are no longer transmitting, in succession. In this example, each civilization emits radiation during a lifetime, $L=10,000$ years, 
creating annuli widths, $cL=10,000$ lyr, $1/5$ the radius of the Galaxy, $R_G=50,000$ lyr. 
The civilizations form at a rate of one every $6000$ years, i.e. $\eta_b=0.000166$ yr$^{-1}$, illustrated by the difference in the radii 
of the successive annuli, $6000$ lyr. The small, light-blue circle at 
lower right represents a civilization that is still emitting EM radiation, having started $7,300$ years ago, and having another $2700$ 
years to go. The other three civilizations with larger outer radii started emitting EM radiation over $10,000$ years ago and have since ceased emitting, leaving 
their radiation annuli. The civilizations that formed before these four are not shown to avoid cluttering the diagram.}\label{fig1}
\end{center}
\end{figure}

By definition, the technological civilizations launch EM waves (of unspecified wavelengths) into the Galaxy, presumably in many 
directions.  We model the emission as isotropic.  To the extent that technological civilizations launch EM waves non-isotropically, 
i.e. in narrow beams, the volume and areas occupied by the radiation will be less than is computed in this paper, lowering the probability of 
detection. The leading edge of the EM waves constitutes a signal ``front'' of notionally spherical shape. When the civilization ceases 
emitting, the last waves similarly leave the planet at the speed of light, constituting a trailing front. The leading and trailing fronts are 
emitted at two times separated by a duration, $L$, producing an electromagnetic ``shell wall'' having thickness, $cL$, where $c$ is the speed 
of light. The quantity, $cL$, is identical to $\Delta$ in \citet{Grimaldi2017} and $L$ is identical to his $\Delta t$. The leading and trailing 
fronts both travel outward at the speed of light, maintaining a constant distance from each other, creating a shell wall of constant thickness 
filled with EM radiation. 

At any instant, the shell wall occupies some fraction of the entire volume of the Milky Way Galaxy. Only within that fraction of the Galaxy can 
the waves be detected (presuming adequate telescopes). Outside that fraction, the waves don't exist, preventing detection. SETI searches on 
Earth can only detect other civilizations whose signal shell wall intersects the Earth at the present time \citep{Grimaldi2017}. For a signal 
longevity, $L$, shorter than the light-crossing time from two opposite edges of the Milky Way Galaxy, $T_M=(2R_G)/c\sim 100,000$ years, the fraction 
will be less than unity. Note that three time scales have emerged, $T_M$, $L$, and $\tau_b$.  

Figure~\ref{fig1} shows a schematic of the Milky Way Galaxy, with EM waves emitted by five civilizations that formed in succession 
separated by a time, $\tau_b$, each having a lifetime $L$.  Each civilization emits waves propagating at the speed of light away from their origin. 
In this example, $L=10,000$ years and $\tau_b=6000$ years.  As a result, the outer radius of each signal front is $6000$ lyr larger than that emitted 
by the next civilization to emerge. A recently formed civilization is located at lower right, shown by a light blue circle of radius $7300$ lyr, 
which will continue to emit waves for another $2700$ years, and thus has a filled circle of EM waves. The other three civilizations with
larger outer radii were born earlier at intervals of $6000$ years, and have since ceased to emit waves.  All have EM shell walls that are annuli 
with width $cL$, as they intersect the plane of the Galaxy.  Figure~\ref{fig1} thus shows the succession of five propagating signals, each larger 
than the previously emitted one by $c\tau_b=6000$ lyr, three having annuli of thickness $cL$, having ceased emitting after $L$ years, resulting 
in central holes. 

Here, we calculate the probability that the Earth is within the emitted shells of light from other civilizations within the Milky 
Way Galaxy, as a function of $L$ and $\eta_b$. We adopt the simplifying assumption that all civilizations remain technological, actively 
emitting EM radiation, for a single lifetime, $L$, in the spirit of the Drake equation. Later, we may allow the lifetimes to assume a distribution 
among the different civilizations. One can imagine many reasons that a civilization might cease its emission of powerful EM waves, 
including the use of more energy-efficient or private forms of communication, actively hiding from the other civilizations in the Galaxy, or 
simply going extinct. Just in the past two decades, the growing use of modern internet repeaters and fiber optics, along with precisely directed 
lasers, suggests that the leakage of electromagnetic radiation from the Earth may cease to increase, if not decrease with time, 
thus indicating that the effective $L$ for SETI purposes could be shorter than the $12,100$ years estimated by \citet{Gott1993}.

Civilizations presumably emerge at various locations within the Milky Way Galaxy at a rate proportional to the stellar density, and more 
specifically to the density of long-lived, habitable planets that permit stable evolution.  We do not know the types of stars nor their many 
properties that are conducive to technological life.  The accurate calculation of the probability that signals are crossing the Earth requires, 
intrinsically, a numerical solution that includes the full 3D shape of the Milky Way Galaxy and the number density of stars as a function 
of position. The density of stars that may harbor technological life provides the ``weighting factor'' for the probability that a civilization is 
born, launching EM waves. We postpone a numerical treatment that utilizes full models of the Milky Way Galaxy, e.g., Besancon \citep{Robin2003} or 
TRILEGAL \citep{Girardi2012}. 
They could be employed to render the emergence of technological life as proportional to the density of various types of stars, e.g. G, K, or M-type 
main sequence stars. For the moment, we must bypass any specificity of the association of stellar types with technological 
life and ignore the 3D density distribution of stars. Instead, we adopt the simplification that such life may emerge anywhere within the disk 
of the Milky Way with equal probability.

\begin{figure*}[t]
\begin{center}
\includegraphics[scale=0.3,clip=true]{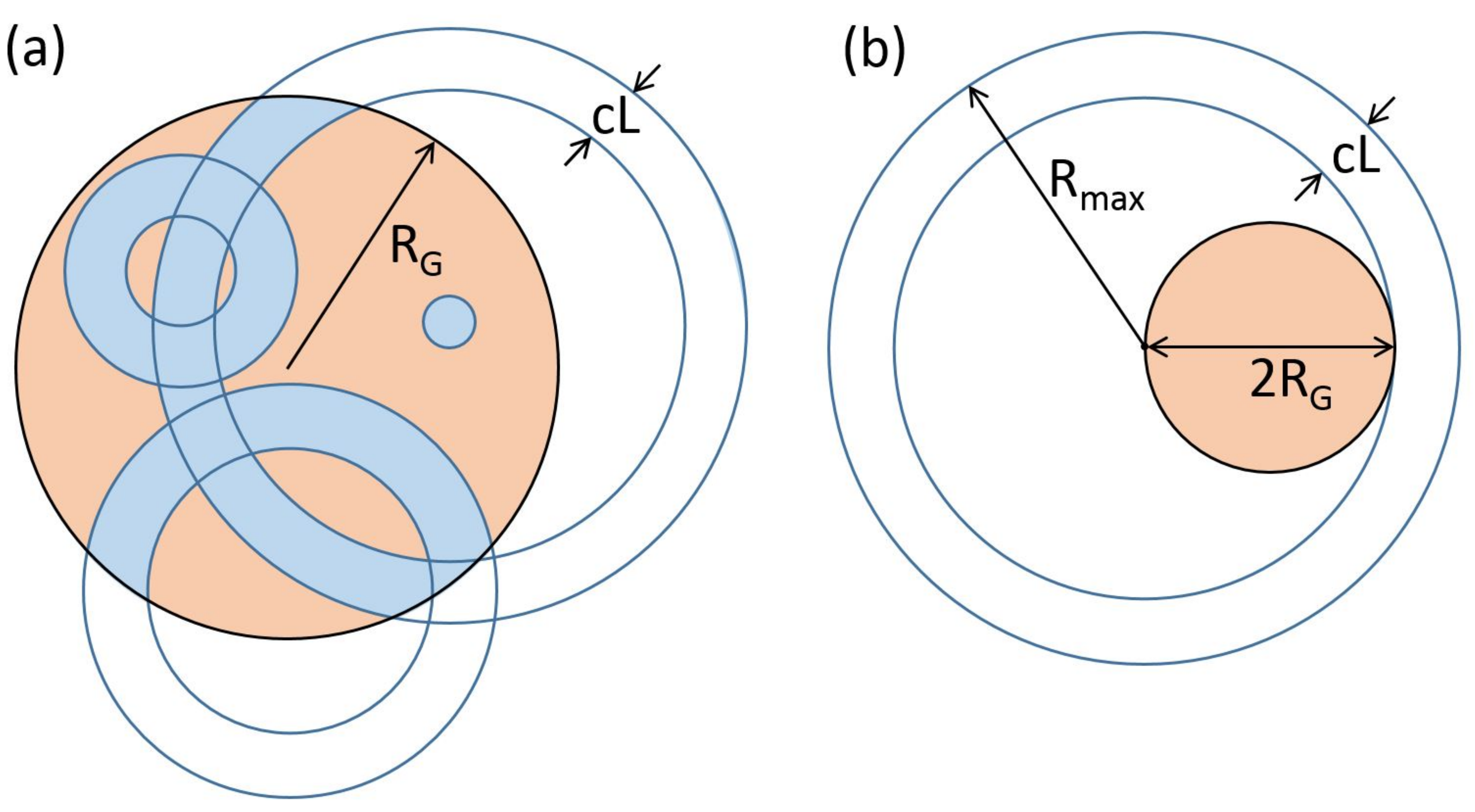}
\caption{(a) Schematic illustration of the galactic disk of radius $R_G$ with annuli representing shell signals having width $cL$ and different outer radii. 
The annuli are centered at random positions within the galaxy. The portions of the annuli depicted in pale blue color highlight the regions of the signals 
intersecting the galactic disk. (b) Since the center of any annulus is located within the galactic disk, there exists for a given cL a maximum outer 
radius, $R_\textrm{max}$, such that annuli with larger outer radii do not intersect the galaxy. The figure shows that $R_\textrm{max}=2R_G+cL$, which 
is obtained by placing the emitting civilization on the edge of the galactic disk and setting the radius of the trailing edge equal to the galactic 
diameter.}\label{fig2}
\end{center}
\end{figure*}

\subsection{Signal Emission Propagation}
\label{emi}

At the instant a civilization becomes ``technological'' it begins transmitting electromagnetic radiation outward. We adopt the simple case of a radiation 
pattern that is approximately spherical, expanding at the speed of light. The leading edge of the radiation has a radius,
\begin{equation}
\label{Rt1} 
R(t)=ct,
\end{equation}
where $t$ is the time elapsed since the first emission.  At some time, $L$, after the first emission that civilization ceases to transmit, causing a trailing 
edge of the shell with radius 
\begin{equation}
\label{Rt2}
R_\textrm{trail}(t)=R(t) - cL. 
\end{equation}
In this model, at a time $\tau_b$ later, another civilization pops into existence, emitting its shell of EM waves.  Additional shells are launched 
every $\tau_b$ years, yielding a population of shells within the Milky Way Galaxy. The spherical shells intersect the Milky Way disk in annuli, each having 
an outer radius, 
\begin{equation}
\label{Rt3} 
R_i(t_i)=ct_i,
\end{equation}
where $t_i$ and $R_i(t_i)$ are the time since emergence, and the current outer radius, of the annulus from the $i$th civilization. Each shell has a leading 
edge that is larger than the leading edge of the next shell emitted by a distance, $c\tau_b$,  the distance traveled by the leading edge of the shell before the 
next one is launched. 

\subsection{Area of the Signals within the Galaxy}
\label{area}

We develop a framework to calculate the number of electromagnetic shell walls that currently intersect a location within the Galaxy, 
such as Earth. \citet{Grimaldi2017} pursues a calculation of the probability that a signal is crossing any point in the Galaxy 
(e.g., Earth), using a probabilistic approach.  Here we calculate a slightly different, but related, geometrical quantity. We estimate 
the area within the Milky Way that is currently covered by EM shell walls from civilizations.  The area may be greater or less than 
the area of the disk of the Milky Way. The ratio of total area of all signal annuli to the total area of the 
Galactic disk offers a measure of the number of civilizations detectable at any location within the Galaxy, on average. For example, 
if the area of all annuli (within the Milky Way disk) is $12\times$ the area of the disk itself, we expect that $12$ signals are typically 
washing over any location within the Milky Way Galaxy at any instant in time. 

We calculate the area of all shells within the disk as follows. We begin by calculating the number of annuli within the Galactic disk. 
Each civilization resides at an arbitrary location within the disk, causing the emitted shell in one azimuthal direction to exit the 
edge of the Galaxy before the portion of the shell on the other side exits. This is illustrated in Fig.~\ref{fig2}a, which highlights 
the portions of the annuli that intersect the galactic disk and are therefore detectable in principle. Conversely, the parts of the 
signals that lie outside the edge of the galaxy are of no interest.  

Since the leading and trailing edges of the annuli grow with time at the speed of light, there is a finite time after which a given 
shell lies completely outside the galactic disk. For example, a civilization located near the central bulge of the Galaxy emits a shell 
signal whose trailing edge takes a time $R_G/c$ to exit the Galaxy. The exit time increases as the location of the civilization moves 
toward the edge of the galactic disk, where it reaches its maximum value, $T_M=2R_G/c$, which is the light travel time from one edge 
of the Galaxy to the diametrically opposite edge, as shown in Fig.~\ref{fig2}b. Therefore, regardless of the position of the emitting 
civilization within the galactic disk, any annulus whose first emission is older than $T_M+L$ resides outside 
the Galaxy, because its inner radius is larger than $2R_G$, and it is thus undetectable. In the following, we shall consider only 
shells that can be in principle detected, and that have thus been emitted since a time $T_M+L$ before present (or, equivalently, 
that have outer radius smaller than $cT_M+cL$, as seen in Fig.~\ref{fig2}b).

We identify each shell with an index, $i$, with $i=1$ representing the smallest (most recently launched) shell. Depending on the time 
of its first emission, this shell has a leading edge with radius $R_1$ comprised between $0$ and $c\tau_b$. The $i=2$ shell has an 
outer radius $R_2$ larger than $R_1$ by $c\tau_b$, and it is therefore comprised between $c\tau_b$ and $2c\tau_b$. We thus assign 
to the $i$th shell a leading edge of radius $R_i  = ct_i$, with $t_i=(i-1)\tau_b+\tau$, where $\tau$ can take any value between 
$\tau=0$ and $\tau=\tau_b$. 

For a given initial time $\tau$, the total number of shells intersecting the Galaxy is computed by counting the number of shells
that have spawned at a time $t_i$ smaller than $T_M+L$. The mean number of such shells, $N_s$, is obtained by averaging over $\tau$: 
\begin{align}
\label{Ns1}
N_s&=\frac{1}{\tau_b}\int_0^{\tau_b}\! d\tau\sum_{i\geq 1}\theta(T_M+L-t_i)\nonumber\\
&=\sum_{i\geq 1}\frac{1}{\tau_b}\int_{(i-1)\tau_b}^{i\tau_b}\!\! dt\,\theta(T_M+L-t)=\frac{T_M+L}{\tau_b},
\end{align}
where $\theta(x)=1$ for $x\geq 0$ and $\theta(x)=0$ for $x<0$, and in the second equality we have changed the variable of integration from 
$\tau$ to $t=(i-1) \tau_b+\tau$.
 
Each shell covers a region of the Galaxy given by the intersection between the annulus centered at the location of the emitting 
civilization and the galactic disk. For a shell emitted at time $t_i$ from an arbitrary point in the Galaxy, the intersection area, 
$A_i$, is given by the difference between the intersection area enclosed by the circle (of radius $R_i$) at the leading edge of the 
shell and the intersection area enclosed by the circle (of radius $R_i-cL$) at the trailing edge. 
The average over the arbitrary location of the center of the annulus within the galactic disk gives: 
\begin{equation}
\label{Ai}
A_i=f(R_i)-f(R_i-cL),
\end{equation}
where
\begin{equation}
\label{effe}
f(\rho)=\frac{1}{\pi R_G^2}\int\! d\vec{r}\int\! d\vec{r}'\,\theta(R_G-r) \theta(R_G-r')\theta(\rho-\vert\vec{r}-\vec{r}'\vert)
\end{equation}
is the average overlap area between the galactic disk and a disk of radius $\rho$. Although Eq.~\eqref{effe} can be calculated exactly 
for any value of $\rho$, we shall make use only of the following self-evident properties:
\begin{equation}
\label{prop}
f(\rho\leq 0)=0,\,\,\,\,\,\,\,\,\, f(\rho\geq 2R_G)=\pi R_G^2.
\end{equation}
Note that for $cL>R_i$ the first of these equalities automatically reduces $A_i$ to the intersecting area for a disk-like signal 
(i.e., having no concentric hole). 

\subsection{Summing the Areas of the Signals within the Galaxy}
\label{summing}

We now sum the areas of the Galaxy covered by all annuli of electromagnetic radiation. We must sum only the intersecting areas 
that result from signals emitted at times $t_i$ smaller than $T_M+L$, that is: $\sum_{i\geq 1} A_i\theta(T_M+L-t_i)$. Using Eq.~\eqref{Ai}, 
the sum of such areas averaged over $\tau$ gives:
\begin{widetext}
\begin{equation}
\label{allshells1}
A_\textrm{all shells}=\frac{1}{\tau_b}\int_0^{\tau_b}\!d\tau\sum_{i\geq 1}A_i\theta(T_M+L-t_i)
=\sum_{i\geq 1}\frac{1}{\tau_b}\int_0^{\tau_b}\!d\tau[f(R_i)-f(R_i-cL)]\theta(T_M+L-t_i)
\end{equation}
which after changing the variable of time integration from $\tau$ to $t=(i-1)\tau_b+\tau$ reduces to
\begin{align}
\label{allshells2}
A_\textrm{all shells}&=\sum_{i\geq 1}\frac{1}{\tau_b}\int_{(i-1)\tau_b}^{i\tau_b}\!dt[f(ct)-f(ct-cL)]\theta(T_M+L-t)
=\frac{1}{\tau_b}\int_0^{T_M+L}\!dt[f(ct)-f(ct-cL)]\nonumber\\
&=\frac{1}{\tau_b}\int_{T_M}^{T_M+L}\!dt\,f(ct),
\end{align}
\end{widetext}
where in the last equality we have used the first identity of Eq.~\eqref{prop}. Since $f(ct)=\pi R_G^2$ for all times larger than $T_M$ 
(second identity of Eq.~\eqref{prop}) we finally obtain:
\begin{equation}
\label{allshells3}
A_\textrm{all shells}=\pi R_G^2\frac{L}{\tau_b}.
\end{equation}
The ratio of the total area of all annuli, $A_\textrm{all shells}$, to the area of the disk of the Galaxy, $A_G$, gives the mean 
number of such annuli that are currently washing over an arbitrary location of the galactic disk. Since $A_G$ is simply $\pi R_G^2$, 
from Eq.~\eqref{allshells3} we find
\begin{equation}
\label{allshells4}
\frac{A_\textrm{all shells}}{A_G}=\frac{L}{\tau_b}.
\end{equation}
The right-hand side of this equation represents the number of galactic civilizations born during the lifetime, $L$, of a 
typical civilization, and it is a quantity that can be arbitrarily large or small (albeit positive).  For $L > \tau_b$, civilizations 
live long enough that other civilizations are born during their lifetime, filling the galaxy with the areas of many annuli of signals. 
In that case, more than one civilization is typically detectable at any time. 
If $L< \tau_b$, the annuli have widths narrower than their difference in radius so that less than one annulus in average washes 
over an arbitrary point in the Galaxy. 

By using Eq.~\eqref{Ns1}, we can represent the mean number of detectable signals in terms of the mean number $N_s$ of annuli 
filling the Galaxy:
\begin{equation}
\label{allshells5}
\frac{A_\textrm{all shells}}{A_G} =\frac{L}{T_M+L} N_s,
\end{equation}
where the ratio of the duration of the signal over the age of the oldest detectable annulus, $L/(T_M+L)$, is the scaled signal 
longevity. In \citet{Grimaldi2017} this quantity was equivalently expressed as the ratio between the thickness of the shell and the 
larger outer radius of a detectable signal. The scaled longevity allows us to discuss two important special cases, $L<T_M$ and $L>T_M$, 
in relation to the mean number of signals that can be possibly detected.

\subsection{Important Special Case: $L < T_M$}
\label{imp1}

We consider here the important special case of civilization lifetimes shorter than the light-travel time across the Milky Way Galaxy, $L < T_M$. This case 
is important for at least two reasons. The resulting EM waves fill only an annulus within the Galaxy, never filling it, before exiting. Also, 
the civilizations with lifetimes of thousands of years may be recognizable by current humans. But technological civilizations millions of years old may 
be unrecognizable to us, transformed by inventions and physics beyond our imagination. As an example of $L < T_M$, we may imagine civilizations 
having $L=10,000$ years, i.e., $L\simeq T_M/10$.  They emit EM signals that occupy annuli having width $cL=10,000$ lyr, and ever expanding radii. 
In this example, we see from Eq.~\eqref{allshells5} that the mean number of detectable annuli is about $10$ times smaller than the number of annuli, $N_s$, 
filling the Galaxy. Consequently, the requirement to detect at least one signal from the Galaxy is that $N_s$ must be larger than about $10$.

\subsection{Converse Special Case:  $L> T_M$}
\label{imp2}

We now consider the opposite special case, $L > T_M$, civilizations with lifetimes greater than the light-crossing time, $T_M\approx 100,000$ years, 
of the Galaxy. In this case the signals overfill the Galaxy and cover the entire area of the Galactic disk during their lifetime. 
Yet, the ``extra lifetime'' provides no additional area coverage of the signal compared to the case $L < T_M$ because the signals that propagate 
beyond the radius of the Galaxy offer no additional opportunity for detection by denizens of the Galaxy, e.g. at Earth. This is well described 
by the scaled longevity of the signal, $L/(T_M+L)$, which saturates at unity as $L \gg T_M$.

For $L > \tau_b$ civilizations live longer than the wait time for other civilizations to be born, filling the galaxy with overlapping 
signals from each of them. If $L < \tau_b$, (yet still with $L > T_M$), the signal from each civilization fills the Galaxy, but there is a 
gap in time between when one civilization dies and the next one arises, leaving typically only $N_s<1+T_M/L$ signals within the 
Galaxy at any time.  At any location, e.g. Earth, there is at 
most one signal washing over it at any instant, so that the probability of a signal washing over the Earth is less than unity.  

\subsection{Relation to Drake's $N$ }
\label{drake}

As before, we define Drake's $N$ as the number of civilizations currently emitting EM waves, and their rate of emergence in the Galaxy as 
$\eta_b=1/\tau_b$. The transmitting civilizations notionally spring from the reservoir of stars and their planets in the Milky Way Galaxy. 
Note that a given planet could have a civilization emerge and die, leaving it available for another species to emerge at a later time, 
incorporated into $\tau_b$. In our simple model, $L$ (measured in years) has essentially the same meaning as that in the classic Drake 
equation.  Each year, a fraction, $1/L$, of the existing civilizations ceases its emission of EM waves.  If $N$ is the number of civilizations 
at any time, they have a spread of ages dictated by their birthrate, some being nearly $L$ years old and soon to cease emissions, constituting 
a civilization conveyor belt of length, $L$, and item separation, $\tau_b$. The number of them that cease emitting per year is $N/L$. 
For example, if $N=10,000$ and $L=5000$ years , two civilizations cease emitting every year so that after $5000$ years all 
$10,000$ have ceased, and are replaced, in equilibrium by newly born civilizations.

To solve for equilibrium we equate a birth rate to the death rate:
\begin{equation}
\label{drake1}
\eta_b=\frac{1}{\tau_b}=\frac{N}{L}.
\end{equation}
We take the death rate to be linear with time, $N/L$, not exponential.  Note that equilibrium must be reached eventually in the life 
of a galaxy after sufficient heavy elements form, which happens within the first billion years. In the early days of a galaxy, the number 
of civilizations ramps up until finally the number of them is so great that the death rate matches the fixed birth rate. This situation is 
different from a population of a biological species, in which the birth rate depends on the number of existing members of that population. 
Here, the birth rate of new civilizations does not depend on $N$. Thus, we have, from 
Eq.~\eqref{drake1}:
\begin{equation}
\label{drake2}
N=\frac{L}{\tau_b}.
\end{equation}
This is Drake's $N$ in terms of the quantities in the present formalism.  The relation is obvious in hindsight as the entire lifetime 
of any civilization is simply chopped up into temporal segments of length $\tau_b$ during which a cohort of civilizations, numbering 
$L/\tau_b$, arise. Indeed, Eq.~\eqref{drake2} simply is the Drake equation itself in which the right hand side has been compartmentalized 
into two terms, one being the usual $L$ and the other is $1/\tau_b$ representing all the other terms, i.e. the rate of formation of 
technological, communicating civilizations. This version of the Drake equation is also found 
by \citet{Balbi2017} who relates it to Little's Law \citep{Little1961}.

Comparison of Eq.~\eqref{drake2}  with Eq.~\eqref{allshells4} leads us to the following 
equivalence:
\begin{equation}
\label{drake3}
\frac{A_\textrm{all shells}}{A_G}=N,
\end{equation}
that is, the ratio of the area of all signals within the Galaxy to the area of the galactic disk (or, equivalently, the mean number 
of detectable annuli) is approximately equal to the number of civilizations currently emitting.

This result may seem surprising at first, but it can be interpreted conceptually by examining the case of $N=1$, for which just one 
civilization is currently 
emitting a signal annulus at any time. If the duration of the signal is much smaller than the light-time crossing of the galaxy, 
$L\ll T_M$, one wonders how the typical number of detectable annuli could be still unity, as predicted by the Eq.~\eqref{drake3}.  
The interpretation is that $N=1$ implies that the 
galaxy typically has one civilization emitting signals at all times, implying that despite its short lifetime compared to $T_M$, 
previous civilizations had been emitting signals allowing that cohort to pave the entire area of the plane of the Galaxy with signals. 
This is readily seen from Eq.~\eqref{allshells5}, from which we obtain at once that, for $N=1$ and $L\ll T_M$, the number of the annuli 
filling the Galaxy is $N_s>T_M/L\gg 1$, meaning that beside the signal that is currently emitted, there are other signal annuli in the 
Galaxy from civilizations that are now extinct. Note that the civilization that exists at the present 
time is probably not the one whose signal is currently washing over the Earth. 

Let us now examine the case in which the currently transmitting civilization ($N=1$) emits a signal for a time $L$ much larger than $T_M$. 
In that case, the civilization transmits for a time, $L$, filling the Galaxy with its signal, at which point the civilization dies, but 
another is immediately born, quickly filling 
the Galaxy with its signal behind the exiting signal of the previous one. Note the leading edge of the new signal emerges just as the 
trailing edge of the previous signal is just leaving its transmitter. Thus the fractional areas of the two signals add up to the area of 
the whole Galaxy, leaving the Galaxy essentially filled at all times.

Equation \eqref{drake3} implies that a typical patch of the disk of the galaxy has a number, $N$, of electromagnetic signals coincident with that patch,
where $N$ is also the number of currently transmitting civilizations in the Galaxy (including those whose signals havenÕt arrived at the patch). 
The expected number of signals washing over Earth (or any point in the Galaxy) is equal to the equilibrium (typical) number of civilizations actively 
transmitting in the Galaxy, $N$. It worth stressing, however, that this equivalence holds true only for the case in which the transmitted signals are 
isotropic. For example, highly anisotropic, beam-like signals that are directed in random directions in space have a much lower probability to hit an 
arbitrary location in the Galaxy \citep{Forgan2014, Grimaldi2017}. Inclusion of any such beaming would obviously reduce the ratio of the area of the 
signals to the area of the Galaxy. In this case, the fractional area filled by the signals, $A_\textrm{all shells}/A_G$, is reduced by a factor
proportional to the solid angle of the beams.  
 
To summarize, isotropic signals emitted from each civilization intersect the Galactic disk in annuli. But the mean number of annuli
intersecting the Earth is simply Drake's $N$. 

\section{Discussion}
\label{discuss}

\subsection{Implications for SETI and Future Improvements to the Model}
\label{disc1}
Our simple model with arguments regarding the birth and death of civilizations has demonstrated that the ratio of the total area covered by EM 
signals to the area of the Galactic disk is simply the current number of active technological civilizations, $N$, currently in the Galaxy.  
Notably, the narrow annuli of EM radiation emitted by civilizations that live much less than $\sim 100,000$ years individually fill only a small 
fraction of the area of the Galaxy,  
But the sum of their areas is $N$ times the area of the Galaxy itself, where $N$ is the number of actively emitting civilizations.  
Since $N = L/\tau_b$, the unknown birthrate and lifetime of civilizations leaves unknown the number, $N$, of different EM signals crossing 
Earth. However, the transmissions arriving at Earth may come from distant civilizations that ceased transmitting long ago, while many 
civilizations still actively emitting may have sent signals that have not arrived yet. 

In this paper, we adopted the simplifying approximation of single values for the characteristic lifetime and birthrate, $L$ and $\tau_b$, of detectable 
transmitters. However, as long as we assume that the durations of the signals are independent of the time of their first emission, averages over their 
distributions can be performed to compute $N$ and the number of signals at Earth. Indeed, since the mean number of annuli, $N_s$, and the total area 
occupied by the signals in the Galaxy, $A_\textrm{all signals}$, both depend linearly on the lifetime of the signal, we can replace $L$ in 
Eqs.~\eqref{Ns1}, \eqref{allshells3} and \eqref{allshells4} by its average, $\langle L\rangle$, over the corresponding distribution of the 
signal duration. 
Similarly, if the time intervals between any two consecutively spawned shells are identically distributed we can replace 
$\tau_b$ by $1/\langle\tau_b^{ -1}\rangle$ in Eqs.~\eqref{Ns1} and \eqref{allshells4}. The equivalence between 
Drake's $N$ and the mean number of detectable signals, Eq.~\eqref{drake3}, remains unaffected.

In this paper, we derive simple scaling functions based on the three time scales, $L$, $\tau_b$, and the light-crossing time of the Milky Way 
Galaxy, $T_M$. 
This is overly simplistic, because a luminosity function of ETIs that remains non-zero at extremely high luminosity, over 
$100,000$ TW, while rare, may imply that sources located in other galaxies offer the highest probability of detection.  
If so, the formalism presented here should be recast to involve intergalactic 
light-crossing times to the nearest luminous ETI source, along with a luminosity-weighted average $L$, and associated $\tau_b$. For example, if the 
nearest detectable civilization is located in another galaxy within the local group, the interpretations of a typical $L$, $\tau_b$, and the light-crossing time of 
the Local Group should be employed. This issue highlights a general concern that the optimal targets for a SETI search depend sensitively on the unknown 
luminosity function of ETIs. Indeed, if very high luminosity emission is the norm, and $N$ is small ($<1$), then looking outside of the Milky Way may 
improve our probability of detecting intelligent life. Also, if extremely high luminosity occurs, the signal may be detected from 
such emitters anywhere in the Galaxy. But they will be rare. Best to search in directions where the column density of stars is greatest; that is, in 
the galactic plane at small galactic longitudes.

The approximate values of $L$ and $\tau_b$ in the Milky Way Galaxy remain utterly unknown.  Still, based on the spawning and lifetimes of 
other species on Earth, some have ventured estimates within orders of magnitude.  \citet{Gott1993} provides a 95\% confidence level upper 
limit on the mean longevity of 
radio-transmitting civilizations, giving, $L<12,000$ years. He also estimates the rate, $1/\tau_b$ , at which civilizations are forming in the Galaxy to 
be less than $0.01$ per year, i.e. $\tau_b>100$ years. Combining these estimates yields $N=L/\tau_b<120$, which is also the approximate upper limit of the number of transmissions crossing an arbitrary point in the Galaxy. That is, according to this estimate, SETI should detect no more than $120$ 
different signals, a bounty compared to the present zero. 

Beyond the scope of this paper but certainly important is the possibility that a single civilization may multiply its emission of EM 
radiation by migration and colonization or by robotic (and perhaps self-replicating) probes, all of which could maintain transmitters of 
radiation \citep{Gertz2017}. Colonization and probes imply that the lifetime, $L$, is related to the effective birthrate of new civilizations 
(or probes), augmenting $\eta_b$, rendering the two parameters mutually dependent. A proliferation of colonies and probes could be the dominant 
source of EM signals from ``technological civilizations'' in the Galaxy. Modeling such proliferation involves parametrization of the likelihood 
that a civilization can bodily migrate to new outposts, either within its planetary system or to neighboring stars, and modeling the probability 
and proliferation due to subsequent migrations, increasing exponentially the number of lineages like a genealogical tree. 

The changing luminosity of host stars may motivate such migrations \citep{Zuckerman1985, Zuckerman1995}. However, \citet{Gott1995} has argued 
that migration of civilizations may be insignificant, otherwise we would likely be one of the migrants. Moreover, migration over interstellar distances 
is easily shown to be extraordinarily difficult from energy needs, centuries needed for transport, and technical challenges. Still, the proliferation of 
robotic probes that emit intense EM radiation for communication remains plausible. If we define the birthrate of ``civilizations'', $\eta_b$, 
to include robotic offspring that emit EM radiation across interstellar distances, the coupling to $L$ may change significantly the probability 
of signals crossing the Earth relative to Drake's $N$.  One may need to redefine Drake's $N$ to include all transmitting robots, as they take the 
role of independent civilizations.

\subsection{Implications for $L$ and Earth's Visibility}
\label{disc2}

In this paper, the value of $L$ is defined  as the duration of detectable EM transmissions emitted by civilizations.  However this definition fails 
to account for the changes with time of detectable transmitted power.  Thus each civilization  will arrive at a different value of 
observable $L$,  strongly influenced by their development of high-powered transmitters and sensitive receivers, and possible invention of novel 
observing systems such as using their star as a gravitational lens.
Similarly, the EM power from a civilization may vary by orders of magnitude in its spectral energy distribution and angular distribution over its sky.
Surely, a useful definition of  $L$ includes some threshold flux for detectability.

Optical and infrared SETI brings its own added set of cautionary issues regarding the formalism in this paper 
\citep{Werthimer2001, Ekers2002, Howard2004}. 
A current best Optical/IR laser launcher has an aperture of $10$ meters, yielding a diffraction-limited beam with an opening angle of 
$\sim 0.02$ arcsec. Such a small angle has the advantage of pouring all of the laser power into a small, AU-scale footprint depending on 
distance from Earth. But if the signal is to be intercepted by the Earth, the 
transmitter must be aimed with high pointing precision not at the Earth but at where the Earth will be when the laser beam 
arrives, e.g. \citep{Tellis2017}.  
Thus, the definition of $L$ pertains to optical/IR SETI with the additional constraint that civilizations either intentionally are able to 
aim a narrow beacon 
with great precision and foresight, or that they are emitting large numbers of laser beams in many carefully chosen directions, creating some sizable 
probability that one of the beams intercepts Earth serendipitously.  Thus, the lifetime of active transmissions, $L$, for optical/IR laser beams requires not 
just a power threshold but an additional ``persistence lifetime'' regarding either a civilization's continuing effort to signal Earth or their continued 
proliferation of a multitude of powerful lasers.

Of course a civilization would be more likely to intentionally transmit beacons of EM radiation toward Earth if it 
sensed the existence of recipients here. 
Our radio EM transmissions could provide them  with evidence of our existence, providing a rationale for transmitting beams back toward Earth 
\citep{Brin2014, Shostak2014, Zaitsev2014, Gertz2016, Vakoch2016}.  But, our radio EM transmission ``donuts'' have
thickness of only $\sim100$ lyr at present, implying that hypothetical extraterrestrial transmitters must reside 
within $50$ lyr for there to have been time to receive and send back a signal that reaches Earth in the near future. 
However, the intensities of our leakage radio and optical radiation are typically low.   

More powerful than typical terrestrial radio leakage is radar power sent from the Arecibo Observatory with its terawatt-class 
effective radiated power. 
Arecibo has been beaming in various directions since the late 1960s, and with higher power after an upgrade in the late 1990s.  Arecibo 
radar is used to characterize asteroids, study the subsurface structure of terrestrial planets and moons, and probe the Earth's ionosphere. The $2.4$ GHz 
Arecibo radar has a $2$ arcmin beam width  and an effective radiated power (EIRP) of $20$ TW.  At $430$ MHz ($2.5$ TW EIRP) and $47$ MHz ($0.3$ TW EIRP) the 
beams have inversely larger beams but lower EIRP. 
The Arecibo transmission at 2.4 GHz can be detected 50,000 lyr away given a sufficiently large receiver, though diluted by distance, line broadening, and scintillation from the interstellar medium \citep{Cordes1997, Enriquez2017}. Of course, the Arecibo radar signal won't 
arrive there for $50,000$ years, but each transmission beam will intercept many stars along the way, depending on the galactic 
latitude of the transmission. In comparison, the ballistic missile early warning radar systems of various militaries, operate at only MW power levels.  

The Arecibo radar waves propagate toward stars that fortuitously reside within the primary beam and sidelobes.  
The typical Arecibo radar emission used to study asteroids contains a continuous, narrow band signal lasting for many minutes, 
followed by a broadband pseudo-random sequence of pulses that enables to measure light-travel time (distance) and Doppler shifts [see \citep{Naidu2015}].
Thus, Arecibo has been inadvertently sending thousands of powerful radio signals into the Galaxy easily identified as 
of intelligent origin. In addition, radar studies of the Earth's ionosphere have similarly involved the full power (at lower frequencies) 
of Arecibo radar. The asteroid study transmissions have comparable power to the famous Arecibo radar transmission toward the globular star 
cluster M13 that lasted $168$ s in 1974. The large number of stars which have been within all of those 
Arecibo radar transmission beams will receive a signal roughly a million times the intensity of the major 
leakage of the typical non-radar signals from Earth, effectively constituting messages to extraterrestrial intelligences (METI). 
Thus, METI has already happened, starting decades ago, at no cost, and with all the power humanity can now muster. A large 
fraction of the hypothetical technological civilizations residing within a few degrees of our ecliptic plane will receive 
strong signals from Arecibo. The thousands of 
Arecibo radar transmissions constitute the most powerful evidence sent into the Galaxy of the existence of humankind. 

\section{Acknowledgments }

\indent 
This work benefited greatly from discussions with Dan Werthimer, Andrew Siemion, Ben Zuckerman, Jason Wright, Joao Alves, Andrew Fraknoi, John 
Gertz, Avi Loeb, Seth Shostak, Jill Tarter, Shelley Wright, Andrew Howard, Remington Stone, and Paul Horowitz.  We have been inspired and informed 
by the Breakthrough Listen Initiative and by conversations with Yuri Milner, Pete Worden, Pete Klupar, and Jamie Drew.

\end{document}